\documentclass[onecolumn,showpacs,floatfix,nofootinbib,aps]{revtex4}[11pt]
\usepackage{amssymb,amsmath}
\usepackage{subfigure}
\usepackage[dvips]{graphics,color}
\usepackage{epsfig}\usepackage{float}

\newcommand{\n}{\nonumber\\}

\newcommand{\bec}{\begin{center}}
\newcommand{\eec}{\end{center}}

\newcommand{\bea}{\begin{array}}
\newcommand{\ear}{\end{array}}

\newcommand{\II}{\mathbb{I}}
\newcommand{\OO}{\mathbb{O}}

\newcommand{\bfr}{\begin{flushright}}

\newcommand{\efr}{\end{flushright}}
\newcommand{\noi}{\noindent}
\newcommand{\me}{\frac{1}{2}}

\newcommand{\cl}{{\mt{C}}\ell}

\newcommand{\RR}{\mathbb{R}}

\newcommand{\ot}{\otimes}
\newcommand{\la}{\Lambda}
\newcommand{\w}{\wedge}
\newcommand{\g}{\gamma}

\newcommand{\beq}{\begin{eqnarray}}\newcommand{\benu}{\begin{enumerate}}\newcommand{\enu}{\end{enumerate}}
\newcommand{\eeq}{\end{eqnarray}}
\newcommand{\mt}{\mathcal}

\newcommand{\pa}{\partial}

\newcommand{\CC}{\mathbb{C}}

\newcommand{\mk}{\mathfrak}

\newcommand{\bx}{\begin{pmatrix}}
\newcommand{\ex}{\end{pmatrix}}
\newcommand{\ka}{\kappa}

\newcommand{\vt}{\vartheta}

\begin{document}

\title{ELKO Spinor Fields:  Lagrangians for Gravity derived from Supergravity}
\author{Rold\~ao da Rocha}
\email{roldao.rocha@ufabc.edu.br} \affiliation{Centro de
Matem\'atica, Computa\c c\~ao e Cogni\c c\~ao, Universidade
Federal do ABC, 09210-170, Santo Andr\'e, SP, Brazil}
\author{J. M. Hoff da Silva}
\email{hoff@ift.unesp.br} \affiliation{Instituto de F\'{\i}sica
Te\'orica, Universidade Estadual Paulista, Rua Pamplona 145
01405-900 S\~ao Paulo, SP, Brazil}

\begin{abstract}
Dual-helicity eigenspinors of the charge conjugation operator
(ELKO spinor fields) belong --- together with Majorana spinor
fields --- to a wider class of spinor fields, the so-called
flagpole spinor fields, corresponding to the class-(5), according
to Lounesto spinor field classification based on the relations and
values taken by their associated bilinear covariants. There exists
only six such disjoint classes: the first three corresponding to
Dirac spinor fields, and the other three respectively
corresponding to flagpole, flag-dipole and Weyl spinor fields. Using 
the mapping from ELKO spinor fields to the three classes Dirac spinor fields, it is
shown that the Einstein-Hilbert, the Einstein-Palatini, and the
Holst actions can be derived from the Quadratic Spinor Lagrangian
(QSL), as the prime Lagrangian for supergravity. The Holst action is related to
the Ashtekar's quantum gravity formulation. To each one of
these classes, there corresponds a unique kind of action for a
covariant gravity theory. Furthermore we consider the necessary and
sufficient conditions to map Dirac spinor fields (DSFs) to ELKO,
in order to naturally extend the Standard Model to spinor fields
possessing mass dimension one. As ELKO is a prime candidate to
describe dark matter and can be obtained from the DSFs, via a
mapping explicitly constructed that does not preserve spinor
field classes, we prove that --- in particular --- the
Einstein-Hilbert, Einstein-Palatini, and Holst actions can be
derived from the QSL, as a fundamental Lagrangian for
supergravity, via ELKO spinor fields. The geometric meaning of the
mass dimension-transmuting operator --- leading ELKO Lagrangian
into the Dirac Lagrangian --- is also pointed out, together with its relationship to the instanton Hopf fibration.
\end{abstract}
\maketitle

\section{Introduction}

Spinor fields can be classified according to the values assumed by
their respective bilinear covariants. There are only six classes
of spinor fields \cite{lou1,lou2,holl}: three of them are related
to the three non-equivalent classes of Dirac spinor fields (DSFs),
and the others are constituted respectively by the so-called
flag-dipole, flagpole and Weyl spinor fields
\cite{lou1,lou2,holl}. Majorana and ELKO (\emph{Eigenspinoren des
Ladungskonjugationsoperators}, or dual-helicity eigenspinors of
the charge conjugation operator) spinor fields are special
subclasses of flagpole spinor fields \cite{ro1}.

ELKO spinor fields are unexpected spin one-half --- presenting
mass dimension 1 --- matter fields, which belong to a non-standard
Wigner class \cite{alu1,alu2}, and are obtained from a complete
set of dual-helicity eigenspinors of the charge conjugation
operator. Due to the unusual mass dimension, ELKO spinor fields
interact in few possibilities with the Standard Model particles,
which instigates it to be a prime candidate to describe dark
matter\footnote{Other motivations for the ELKO to be a prime
candidate to describe dark matter can be seen in, e.g.,
\cite{alu1,alu2,gau}.}. Indeed,  the new matter fields ---
constructed via ELKO \cite{gau} --- are dark with respect to the
matter and gauge fields of the Standard Model (SM), interacting
only with gravity and the Higgs boson \cite{alu1,alu2,gau,osmano,osmano1}.
Moreover, it is essential  to try to incorporate ELKO spinor
fields in some extension of the SM, identifying  new fields to
dark matter and suggesting how  the dark matter sector Lagrangian
density arises from a mass dimension-transmuting symmetry. We
additionally have already considered the possibility of
incorporating the dynamics of ELKO spinor fields, extending the SM
in order to accomplish the dynamical, as well the not less
fundamental, algebraic, topological and geometric properties,
associated with ELKO. In \cite{osmano}  the underlying equivalence
between Dirac spinor fields (DSFs) and ELKO was analyzed and
investigated and the conditions under which the DSFs can be led to
an ELKO were constructed, since they are inherently distinct and
represent disjoint classes in Lounesto spinor field
classification\footnote{For instance, while the latter belongs to
class (5)
 under such classification, the former is a representative of spinor fields of
types-(1), -(2), and -(3). In addition, when acting on ELKO, the
parity $\mathbb{P}$ and charge conjugation $C$ operators
\emph{commutes} and $\mathbb{P}^2 = -1$, while when acting upon
Dirac spinor fields, such operators \emph{anticommutes} and
$\mathbb{P}^2 = 1$. Besides, $C\mathbb{P}\mathbb{T}$ equals $+1$
and $-1$, respectively for DSFs and ELKO.}. Any invertible map that
takes Dirac particles and leads to ELKO is also capable to make
mass dimension transmutations, since DSFs present mass dimension
three-halves, instead of mass dimension one associated with ELKO.
In this previous paper \cite{osmano} \footnote{R. da Rocha thanks
to Prof. Dharamvir Ahluwalia for
 private communication on the subject.}  the initial pre-requisites to construct
 a natural extension of the Standard
Model (SM) in order to incorporate ELKO were provided, and
consequently a possible description of dark matter
\cite{alu1,alu2,alu3,ahlu4,boe1} in this context.

By using one specific class of DSF --- seen as an equivalence
class of Dirac spinor fields --- and imposing a condition of
constant spinor field, it has already been shown that the
Einstein-Hilbert Lagrangian of General Relativity (GR), as well as
the Lagrangian of its teleparallel equivalent (GR$_\|$), can be
recast as a quadratic spinor Lagrangian (QSL)
\cite{tn1,tung1,tung3}. This development was purposed in an
attempt to better understand the question of the gravitational
energy-momentum localization.

In order to prove the equivalence between the QSL and the
Lagrangians associated with GR and GR$_\|$, a DSF of class-(2) ---
under Lounesto spinor field classification --- with constant
coefficients was used in reference \cite{tn1}. Although this is a
natural choice in the context of the QSL formalism of
gravitational theory, it remains to be better justified.
Furthermore, the use of spinor fields with constant coefficients
is quite restrictive. It is true that one can force the DSF to
have constant coefficients. This is possible because both the
orthonormal frame field and the DSF symmetries, under Lorentz
transformations, can be tied together \cite{tn1}. However, there
are many other possible choices that \emph{do not} require the
orthonormal frame gauge freedom to be the same as the DSF gauge
freedom. In these cases, the rules of the Clifford algebra-valued
differential forms imply the existence of extra terms in  the
boundary term associated with the QSL.

In a previous paper \cite{jpe}, the equivalence between the underlying
algebraic structure of the DSFs and the corresponding gravity
theory actions were established, and one of the main aims of the
present paper is to obtain the Lagrangians of some of the current
theories for gravity and quantum gravity exclusively using ELKO
spinor fields. This equivalence enables us to better characterize
and understand the nature of the spinor field that constitutes the
QSL. ELKO spinor fields can be led to DSFs, and the
Einstein-Hilbert, the Einstein-Palatini and the Holst\footnote{The
Holst action is shown to be equivalent to the Ashtekar
 formulation of Quantum Gravity \cite{hor}.} actions can be derived from a QSL, when we consider ELKO spinor fields.
We begin by showing first that the spinor-valued 1-form field
entering the QSL has necessarily to be constructed by a tensor
product between a \emph{Dirac spinor field} and a Clifford
algebra-valued 1-form: no other spinor fields can lead either to
the Holst action, or to the particular cases of Einstein-Hilbert
and Einstein-Palatini actions. These three gravitational actions
correspond respectively to a class-(2), class-(3), and class-(1)
DSFs, which present complete correspondence to ELKO spinor fields.  ELKO spinor fields that are mapped
into classes-(2) and -(3) of DSFs together give the
Einstein-Hilbert and Einstein-Palatini actions, and the ELKO
spinor fields that are mapped into class-(1) DSF gives alone the
complete Holst action, which shows up also in the proof of
gravitational theory as a SUSY gauge theory \cite{tung2}.
Furthermore, we assume a more general approach, where the ELKO
spinor field is not a constant spinor field anymore. As a
consequence, the boundary term of the QSL will have many
additional terms that can be related to some physical identities,
and may unravel additional properties.

The paper is organized as follows: after presenting some algebraic
preliminaries in Section \ref{w2}, we briefly introduce in Section
\ref{elko} the ELKO spinor fields as well as we recall the
conditions under which a DSFs can be mapped into an ELKO spinor
field. We also point out the geometric meaning of the mass
dimension-transmuting symmetry between ELKO and Dirac spinor
fields. In Section \ref{w3} we investigate the QSL, and in Section \ref{ui}, after briefly
presenting the Lounesto spinor field classification, as well as
some important features of each spinor field class, we show that
Einstein-Hilbert, Einstein-Palatini, and Holst actions can be
derived from a QSL provided we do not restrict ourselves to the
case of a class-(2) DSF, also deriving such Lagrangians for
gravity, via ELKO spinor fields. In the last Section all the
results obtained are discussed.

\section{Preliminaries}
\label{w2} This section is devoted to briefly introduce the
mathematical pre-requisites to completely recall the definition of
Clifford algebra-valued differential forms on a manifold $M$. For
more details, see, e.g., \cite{moro,ro4}.

We denote by $\mathcal{M=} (M, g,\nabla,\tau_g,\uparrow)$ the
spacetime structure: $M$ denotes a 4-dimensional manifold, $g
\in\sec T_{0}^{2}M$ is the metric associated with the cotangent
bundle, $\nabla$ is the Levi-Civita connection of $ g$,
$\tau_g\in\sec {\displaystyle\Lambda^{4}} (T^{\ast}M)$ defines a
spacetime orientation and $\uparrow$ refers to an equivalence
class of timelike 1-form fields defining a time orientation.  By
$F(M)$ we mean the (principal) bundle of frames, by $\mathbf{P}%
_{\mathrm{SO}_{1,3}^{e}}(M\mathbf{)}$ the orthonormal frame
bundle, and
$P_{\mathrm{SO}_{1,3}^{e}}(M)$ denotes the orthonormal coframe bundle. We consider $M$ a spin manifold, and then there exists $\mathbf{P}_{\mathrm{Spin}%
_{1,3}^{e}}(M\mathbf{)}$ and
$P_{\mathrm{Spin}_{1,3}^{e}}(M\mathbf{)}$ which are respectively
the spin frame and the spin coframe bundles. We denote by
$s:P_{\mathrm{Spin}_{1,3}^{e}}(M\mathbf{)\rightarrow}P_{\mathrm{SO}_{1,3}^{e}%
}(M\mathbf{)}$ the fundamental mapping present in the definition
of $P_{\mathrm{Spin}_{1,3}^{e}}(M\mathbf{)}$. A spin structure on
$M$ consists of a principal fiber bundle
$\mathbf{\pi}_{s}:P_{\mathrm{Spin}_{1,3}^{e}}(M)\rightarrow M$,
with group $\mathrm{Spin}_{1,3}^{e}$, and the map
\begin{equation}
s:P_{\mathrm{Spin}_{1,3}^{e}}(M)\rightarrow P_{\mathrm{SO}_{1,3}^{e}%
}(M)\label{spinor bundle 1}%
\end{equation}
satisfying the following conditions:

(i) $\mathbf{\pi}(s(p))=\mathbf{\pi}_{s}(p),\ \forall p\in P_{\mathrm{Spin}%
_{1,3}^{e}}(M);$ $\pi$ is the projection map of the bundle $P_{\mathrm{SO}%
_{1,3}^{e}}(M)$.

(ii) $s(p \phi)=s(p)\mathrm{Ad}_{\phi},\;\forall p\in
P_{\mathrm{Spin}_{1,3}^{e}}(M)$ and
$\mathrm{Ad}:\mathrm{Spin}_{1,3}^{e}\rightarrow\mathrm{Aut}(\cl_{1,3}),$
$\mathrm{Ad}_{\phi}:\cl_{1,3}\ni \Xi\mapsto
\phi\Xi\phi^{-1}\in\cl_{1,3}$ \cite{moro}.

We recall now that sections of $P_{\mathrm{SO}_{1,3}^{e}%
}(M\mathbf{)}$ are orthonormal coframes, and that sections of $P_{\mathrm{Spin}%
_{1,3}^{e}}(M\mathbf{)}$ are also orthonormal coframes such that
two coframes differing by a $2\pi$ rotation are distinct and two
coframes differing by a $4\pi$ rotation are identified. Next we
introduce the Clifford bundle of differential forms
$\mathcal{C\ell
(}M,g)$, which is a vector bundle associated with $P_{\mathrm{Spin}%
_{1,3}^{e}}(M\mathbf{)}$. Their sections are sums of
non-homogeneous differential forms, which will be called Clifford
fields. We recall that \
$\mathcal{C\ell(}M,g)=P_{\mathrm{SO}_{1,3}^{e}}(M)\times
_{\mathrm{Ad}^{\prime}}\cl_{1,3}$, where $\cl_{1,3}%
\simeq$ M(2,${\mathbb{H}})$ is the spacetime algebra. Details of
the bundle structure are as follows \cite{dimakis,dimakis1,est}:

(1) Let $\mathbf{\pi}_{c}:\mathcal{C}\ell(M,g)\rightarrow M$ be
the canonical projection of $\mathcal{C}\ell(M,g)$ and let
$\{U_{\alpha}\}$ be an open covering of $M$. There are
trivialization mappings
$\mathbf{\psi}_{i}:\mathbf{\pi}_{c}^{-1}(U_{i})\rightarrow U_{i}%
\times\cl_{1,3}$ of the form $\mathbf{\psi}_{i}(p)=(\mathbf{\pi}%
_{c}(p),\psi_{i,x}(p))=(x,\psi_{i,x}(p))$. If $x\in U_{i}\cap
U_{j}$ and $p\in\mathbf{\pi}_{c}^{-1}(x)$, then
\begin{equation}
\psi_{i,x}(p)=h_{ij}(x)\psi_{j,x}(p)
\end{equation}
for $h_{ij}(x)\in\mathrm{Aut}(\cl_{1,3})$, where $h_{ij}:U_{i}\cap
U_{j}\rightarrow\mathrm{Aut}(\cl_{1,3})$ are the transition
mappings of $\mathcal{C}\ell(M,g)$. We recall that every
automorphism of $\cl_{1,3}$ is \textit{inner. }Then,
\begin{equation}
h_{ij}(x)\psi_{j,x}(p)=a_{ij}(x)\psi_{i,x}(p)a_{ij}(x)^{-1} \label{4.4}%
\end{equation}
for some $a_{ij}(x)\in\cl_{1,3}^{\star}$, the group of invertible
elements of $\cl_{1,3}$.

(2) As it is well known, the group $\mathrm{SO}_{1,3}^{e}$ has a
natural extension in the Clifford algebra $\cl_{1,3}$. Indeed, we
know that
$\cl_{1,3}^{\star}$ (the group of invertible elements of $\cl%
_{1,3}$) acts naturally on $\cl_{1,3}$ as an algebra automorphism
through its adjoint representation. A set of \emph{lifts} of the
transition functions of $\mathcal{C}\ell(M,\mathtt{g})$ is a set
of elements $\{a_{ij}\}\subset$ $\cl_{1,3}^{\star}$ such that, if
\footnote{Recall
that $\mathrm{Spin}_{1,3}^{e}=\{\phi\in\cl_{1,3}^{0}:\phi\tilde{\phi}%
=1\}\simeq\mathrm{SL}(2,\mathbb{C)}$ is the universal covering
group of the
restricted Lorentz group $\mathrm{SO}_{1,3}^{e}$. Notice that $\cl%
_{1,3}^{0}\simeq\cl_{3,0}\simeq$ M(2,$\mathbb{C})$, the even
subalgebra of $\cl_{1,3}$ is the Pauli algebra.}
\begin{eqnarray}
&&\mathrm{Ad} :\phi\mapsto\mathrm{Ad}_{\phi} \nonumber \\
&&\mathrm{Ad}_{\phi}(\Xi) =\phi \Xi\phi^{-1}, \quad \forall
\Xi\in\cl_{1,3},
\end{eqnarray}
then $\mathrm{Ad}_{a_{ij}}=h_{ij}$ in all intersections.

(3) Also $\sigma=\mathrm{Ad}|_{\mathrm{Spin}_{1,3}^{e}}$ defines a
group homeomorphism
$\sigma:\mathrm{Spin}_{1,3}^{e}\rightarrow\mathrm{SO}_{1,3}^{e}$
which is onto with kernel $\mathbb{Z}_{2}$. We have that
Ad$_{-1}=$ identity,
and so $\mathrm{Ad}:\mathrm{Spin}_{1,3}^{e}\rightarrow\mathrm{Aut}%
(\cl_{1,3})$ descends to a representation of
$\mathrm{SO}_{1,3}^{e}$.
Let us call $\mathrm{Ad}^{\prime}$ this representation, i.e., $\mathrm{Ad}%
^{\prime}:\mathrm{SO}_{1,3}^{e}\rightarrow\mathrm{Aut}(\cl_{1,3})$.
Then we can write $\mathrm{Ad}_{\sigma(\phi)}^{\prime}\Xi=\mathrm{Ad}_{\phi}%
\Xi=\phi\Xi\phi^{-1}$.

(4) It is clear that the structure group of the Clifford bundle
$\mathcal{C}\ell(M,g)$ is reducible from $\mathrm{Aut}%
(\cl_{1,3})$ to $\mathrm{SO}_{1,3}^{e}$. The transition maps of
the principal bundle of oriented Lorentz cotetrads $P_{\mathrm{SO}_{1,3}^{e}%
}(M)$ can thus be (through $\mathrm{Ad}^{\prime}$) taken as
transition maps for the Clifford bundle. We then have \cite{lawmi}
\begin{equation}
\mathcal{C}\ell(M,g)=P_{\mathrm{SO}_{1,3}^{e}}(M)\times
_{\mathrm{Ad}^{\prime}}\cl_{1,3},
\end{equation}
i.e., the Clifford bundle is a vector bundle associated with the
principal bundle $P_{\mathrm{SO}_{1,3}^{e}}(M)$ of orthonormal
Lorentz coframes.

Recall that $\mathcal{C}\!\ell(T_{x}^{\ast}M,g_{x})$ is also a
vector space over $\mathbb{R}$ which is isomorphic to the exterior
algebra
$\Lambda (T_{x}^{\ast}M)$ of the cotangent space and $\Lambda (T_{x}^{\ast}M)=%
{\displaystyle\oplus_{k=0}^{4}}
\Lambda{}^{k}(T_{x}^{\ast}M)$, where $\Lambda^{k}(T_{x}^{\ast}M)$
is the $\binom{4}{k}$-dimensional space of $k$-forms over a point
$x$ on $M$. There is a natural embedding \ $\Lambda
(T^{\ast}M)\hookrightarrow$ $\mathcal{C}\ell(M,g)$ \cite{lawmi}
and sections of $\mathcal{C}\!\ell(M,g)$ --- Clifford fields ---
can be represented as a sum of non-homogeneous differential forms.
Let $\{e_{a}\}\in\sec\mathbf{P}_{\mathrm{SO}_{1,3}^{e}}(M)$ (the
orthonormal frame bundle) be a tetrad basis for $TU\subset TM$
(given an open set $U\subset M$). Moreover, let
$\{\vartheta^{a}\}\in\sec P_{\mathrm{SO}_{1,3}^{e}}(M)$.
Then, for each ${a}=0,1,2,3$, ${\vartheta}^{a}%
\in\sec\Lambda^{1}(T^{\ast}M)\hookrightarrow\sec\mathcal{C}\!\ell
(M,g$). We recall next the crucial result \cite{moro,lawmi} that
in a spin manifold we have:
\begin{equation}
\mathcal{C}\ell(M,\mathtt{\eta})=P_{\mathrm{Spin}_{1,3}^{e}}(M)\times
_{\mathrm{Ad}}\cl_{1,3}. \label{1new}%
\end{equation}
Spinor fields are sections of vector bundles associated with  the
principal bundle of spinor coframes. The well known Dirac spinor
fields are sections of the bundle
\begin{equation}
S_{c}(M,\mathtt{\eta})=P_{\mathrm{Spin}_{1,3}^{e}}(M)\times_{\mu_{c}}\mathbb{C}^{4}, \label{4.7}%
\end{equation}
with $\mu_{c}$ the $D^{(1/2,0)}\oplus D^{(0,1/2)}$ representation of $\mathrm{Spin}%
_{1,3}^{e}\cong\mathrm{SL}(2,\mathbb{C})$ in
$\mathrm{End}(\mathbb{C}^{4})$ ~\cite{choquet}.

The orthonormal coframe field $\{\vartheta^a\}\in\sec\la^1(T^*M)$
can be related to the metric $g$ by
$g=\eta_{ab}\vartheta^a\ot\vartheta^b$, with $(\eta_{ab}) = {\rm
diag} (1,-1,-1,-1)$. In other words, $g$ is the metric on $M$
according to which the elements of $\{e_{a}\}$ are orthonormal
vector fields, i.e., $g_{x}(e_{a}|_{x},e_{b}|_{x}) = \eta_{ab}$
for each $x \in M$. We use the Latin alphabet $a, b, c, \ldots =
0, 1, 2, 3$ to denote anholonomic indices related to the tangent
spaces and the Greek alphabet $\mu, \nu, \rho, \ldots = 0, 1, 2,
3$ to denote holonomic spacetime indices. Let $\{x^{\mu}\}$ be
local coordinates in an open set $U\subset M$. Denoting
$\partial_{\mu}=\partial/\partial x^{\mu}$, one can always expand
the coordinate basis $\{\partial_{\mu}\}$ in terms of $\{e_{a}\}$,
\[
\partial_{\mu} = h^{a}{}_{\mu} e_{a}
\]
for certain functions $h^{a}{}_{\mu}$ on $U$. This immediately
yields $g_{\mu\nu} := g(\partial_{\mu},\partial_{\nu}) =
h^{a}{}_{\mu}h^{b}{}_{\nu} \eta_{ab}$. Consider a Minkowski vector
space $V = \RR^{1,3}$, isomorphic (as a vector space) to $T_xM$
and its associated Clifford algebra $\cl_{1,3}$, generated by the
basis $\{\gamma_\mu\}$ and by the relations $\gamma_\mu\gamma_\nu
+ \gamma_\nu\gamma_\mu = 2\eta_{\mu\nu}$. The Clifford product
will be denoted by juxtaposition. Given two arbitrary (in general
non-homogeneous) form fields $\xi, \zeta\in\sec\Lambda(T^*M)$, the
dual Hodge operator $\star:\sec\Lambda^p(T^*M)
\rightarrow\sec\Lambda^{4-p}(T^*M)$ is defined explicitly by
$\xi\w\star\zeta = G(\xi,\eta)$, where $G:
\sec\Lambda(T^*M)\times\sec\Lambda(T^*M) \rightarrow \RR$ denotes
the metric extended to the space of form fields.

The coframe field $\{\vartheta^a\}$ and the metric-compatible
connection 1-form $\omega^{ab}$ are potentials for the curvature
and the torsion, expressed respectively by the structure equations
\begin{equation}\label{2}
\Omega^a_{\;\;b}=d \omega^a_{\;\;b} + \omega^a_{\;\;\rho}\w
\omega^\rho_{\;\;b}\in\sec\la^2(T^*M) \quad \mbox{and} \quad
\Theta^a = d\vartheta^a + \omega^a_{\;\;b}\wedge
\vartheta^b\in\sec\la^2(T^*M).
\end{equation}
The connection coefficients are implicitly given by
$\omega_{ab}=\omega_{abc}\theta^c$, and the torsion can be
decomposed in its irreducible components under the global Lorentz
group as \cite{h2}
\begin{equation}
\Theta^a = {}^{(1)}\Theta^a + {}^{(2)}\Theta^a + {}^{(3)}\Theta^a
\end{equation}
where
\begin{equation}\label{at3}
{}^{(2)}\Theta^a = \frac{1}{3}\vartheta^a\w(\vartheta^b\lrcorner
\Theta_b), \quad  {}^{(3)}\Theta^a =
-\frac{1}{3}\star(\vartheta^a\w \mk{a}),\quad
 {}^{(1)}\Theta^a = \Theta^a -  {}^{(2)}\Theta^a  - {}^{(3)}\Theta^a,
\end{equation}
with $\mk{a} = \star(\Theta_b\w\vartheta^b)$ denoting the axial
1-form associated with the axial torsion ${}^{(3)}\Theta^a$. The
term $\star{\mk{a}}$ is the well known translational Chern-Simmons
3-form field \cite{h1,h2,h3}, whose total derivative
$d\star\mk{a}$ is the Nieh-Yan 4-form field \cite{z1,z2,ni}.

Clifford algebra-valued differential forms (on Minkowski
spacetime) are elements of $\sec \la(T^*M)\ot\cl_{1,3}$. In
particular, Eqs.~(\ref{2}) are written as
\begin{equation}\label{4}
\Omega = d\omega + \omega\w\omega\ \quad \mbox{and} \quad \Theta =
d\vartheta + \omega\w\vartheta + \vartheta\w\omega,
\end{equation}
where \beq\label{3} \vartheta &=& \vartheta^a\ot\g_a,\qquad \omega
= \frac{1}{4}\omega^{ab}\ot\gamma_{ab},\n \Theta &=&
\Theta^a\ot\g_a,\qquad \Omega =
\frac{1}{4}\Omega^{ab}\ot\gamma_{ab}, \eeq with
$\gamma_{ab}=\me(\gamma_a\g_b - \g_b\g_a)$. All operations in the
exterior algebra of differential forms
 are trivially induced on the space of Clifford-valued differential forms.
In particular, given $\phi^a\in\Lambda(V)$,  the total derivative
$d(\phi^a\ot\g_a)$ is given by $d(\phi^a)\ot\g_a$ and, given a
$p$-form field basis $\{\vt^I\}$ and a Clifford algebra basis
$\{\g_I = \g_a\g_b\g_c\ldots\}$, the exterior product between two
elements $\Phi = \Phi^I\ot\g_I$ and $\Gamma = \Gamma^J\ot\g_J$ of
$\sec\la(T^*M)\ot\cl_{1,3}$ is given by \cite{dimakis,est}
\begin{equation}
\Phi\w\Gamma = (\Phi^I\ot\g_I)\w(\Gamma^J\ot\g_J) =
(\Phi^I\w\Gamma^J)\ot\g_I\g_J.
\end{equation}

\section{ELKO spinor fields}
\label{elko}

In this Section the formal properties of ELKO spinor fields are
briefly revised \cite{alu1,alu2,alu3} and the map between ELKO
spinor fields and DSFs recalled. An ELKO, denoted by $\Psi$,
corresponding to a plane wave with momentum $p=(p^{0},\mathbf{p)}$
can be written, without loss of generality, as
$\Psi(p)=\lambda({\bf p}) e^{-i{p\cdot x}}$ (or
$\Psi(p)=\lambda({\bf p}) e^{i{p\cdot x}}$) where
\begin{equation}
\lambda({\bf
p})=\binom{i\Phi\phi_{L}^{\ast}(\mathbf{p})}{\phi_{L}(\mathbf{p})},
\label{1}%
\end{equation}
\noindent  $\phi_{L}(\mathbf{p})$ denotes a left-handed Weyl
spinor, and given the rotation generators denoted by
${\mathfrak{J}}$, the Wigner's spin-1/2 time reversal operator
$\Phi$ satisfies $\Phi
\mathfrak{J}\Phi^{-1}=-\mathfrak{J}^{\ast}$. Hereon, as in
\cite{alu1}, the Weyl representation of $\gamma^{\mu}$ is used,
i.e.,
\begin{equation}
\gamma_{0}=\gamma^{0}=%
\begin{pmatrix}
\OO & \II\\
\II & \OO
\end{pmatrix}
,\quad-\gamma_{k}=\gamma^{k}=%
\begin{pmatrix}
\OO & -\sigma_{k}\\
\sigma_{k} & \OO
\end{pmatrix}
,\quad
\gamma^{5}=-i\gamma^{0}\gamma^{1}\gamma^{2}\gamma^{3}=-i\gamma^{0123}=
\begin{pmatrix}
\II&\OO \\
\OO & -\II
\end{pmatrix}
\label{dirac matrices}%
\end{equation}
\noindent where
\begin{equation}
\II= \begin{pmatrix}
1 & 0\\
0 & 1
\end{pmatrix}
,\quad \OO=\begin{pmatrix}
0 & 0\\
0 & 0
\end{pmatrix}
,\quad
\sigma_{1}=%
\begin{pmatrix}
0 & 1\\
1 & 0
\end{pmatrix}
,\quad\sigma_{2}=%
\begin{pmatrix}
0 & -i\\
i & 0
\end{pmatrix}
,\quad\sigma_{3}=%
\begin{pmatrix}
1 & 0\\
0 & -1
\end{pmatrix}.
\end{equation}
\noindent    ELKO spinor fields are eigenspinors of the charge
conjugation operator $C$, i.e., $C\lambda(\bf{p})=\pm \lambda({\bf
p})$, for $
C=%
\begin{pmatrix}
\OO & i\Phi \\
-i\Phi & \OO
\end{pmatrix}\,K.$  The operator $K$ is responsible for the $\mathbb{C}$-conjugation of
spinor fields appearing on the right. The plus sign stands for
{\it self-conjugate} spinors, $\lambda^{S}({\bf p})$, while the
minus yields {\it anti self-conjugate} spinors, $\lambda^{A}({\bf
p})$. Explicitly, the complete form of ELKO spinor fields can be
found by solving the equation of helicity
$(\sigma\cdot\widehat{\bf{p}})\phi^{\pm}(\mathbf{0})=\pm
\phi^{\pm}(\mathbf{0})$ in the rest frame and subsequently
performing a boost, in order  to recover the result for any ${\bf
p}$ \cite{alu1}. Note that the helicity of
$i\Phi[\phi_{L}(\mathbf{p})]^\ast$ is opposed to that of
$\phi_L(\mathbf{p})$, since
$(\sigma\cdot\widehat{\bf{p}})\Phi[\phi_L^{{\pm}}(\mathbf{0})]^\ast=\mp
\Phi[\phi_L^{\ast\pm}(\mathbf{0})]^\ast$. Here
$\widehat{\bf{p}}:={\bf p}/\|{\bf
p}\|=(\sin\theta\cos\phi,\sin\theta\sin\phi,\cos\theta)$. The four
spinor fields are given by
\begin{equation}
\lambda^{S/A}_{\{\mp,\pm \}}({\mathbf
p})=\sqrt{\frac{E+m}{2m}}\Bigg(1\mp \frac{{\bf
p}}{E+m}\Bigg)\lambda^{S/A}_{\{\mp,\pm \}}(\bf{0}),
\label{form}\end{equation}
\begin{equation} \qquad\text{where}\qquad
\lambda_{\{\mp,\pm \}}(\bf{0})=%
\begin{pmatrix}
\pm i \Theta[\phi^{\pm}(\bf{0})]^{*} \\
\phi^{\pm}(\bf{0})
\end{pmatrix}
\label{four}.\end{equation} The phases are adopted so that
    \begin{eqnarray}
    \phi^+({\mathbf{0}})= \sqrt{m}\left(\begin{array}{c}
    \cos(\theta/2) e^{-i \phi/2}\\
    \sin(\theta/2) e^{i \phi/2}
    \end{array}\right),\qquad
    \phi^-({\mathbf{0}}) = \sqrt{m}\left(\begin{array}{c}
    -\sin(\theta/2) e^{-i \phi/2}\\
    \cos(\theta/2) e^{i \phi/2}
    \end{array}\right), \label{2222}
    \end{eqnarray}\noi at rest, and since
$\Theta[\phi^{\pm}(\bf{0})]^{*}$ and $\phi^{\pm}(\bf{0})$ present
opposite helicities, ELKO cannot be an eigenspinor field of the
helicity operator, and indeed carries both helicities. In order to
guarantee an invariant real norm, as well as positive definite
norm for two ELKO spinor fields, and negative definite norm for
the other two, the ELKO dual is given by \cite{alu1}
\begin{equation}
\overset{\neg}{\lambda}^{S/A}_{\{\mp,\pm \}}({\bf p})=\pm i \Big[
\lambda^{S/A}_{\{\pm,\mp \}}({\bf p})\Big]^{\dag}\gamma^{0}
\label{dual}.\end{equation}\noindent
 It is useful to
choose $i\Theta=\sigma_{2}$, as in \cite{alu1}, in such a way that
it is possible to express
\begin{equation}
\lambda(\mathbf{p})=\binom{\sigma_{2}\phi_{L}^{\ast}(\mathbf{p})}{\phi_{L}(\mathbf{p})}.
\label{01}%
\end{equation}

Now, any flagpole spinor field is an eigenspinor field of the
charge conjugation operator \cite{lou1,lou2}, here represented by
$\mathcal{C}\psi=-\gamma ^{2}\psi^{\ast}$. Indeed
\begin{align}
-\gamma^{2}\lambda^{\ast}  &  =%
\begin{pmatrix}
0 & \sigma_{2}\\
-\sigma_{2} & 0
\end{pmatrix}
\binom{(\sigma_{2}\phi^{\ast})^{\ast}}{\phi^{\ast}}
=\binom{\sigma_{2}\phi^{\ast}}{-\sigma_{2}\sigma_{2}^{\ast}\phi}\nonumber
=\lambda.
\end{align}
\noindent

Let us make a brief recall of which are the conditions a Dirac
spinor field must obey to be led to an ELKO. In \cite{osmano}
there has been proved that not all DSFs can be led to ELKO, but
only a subset of the three classes
--- under Lounesto classification --- of DSFs restricted to some
conditions. Explicitly, by taking a DSF
\begin{equation}
\psi({\bf p}) = \begin{pmatrix}
\phi_{R}({\bf p})\\
\phi_{L}({\bf p})
\end{pmatrix}
=%
\begin{pmatrix}
\epsilon \sigma_{2}\phi^{*}_{L}({\bf p})\\
\phi_{L}({\bf p})
\end{pmatrix}
\label{comeco},\end{equation} and taking into account that
$\phi_{R}({\bf p})=\chi \phi_{L}({\bf p})$, where $\chi = \frac{E
+ {\mathbf{\sigma}}\cdot {\mathbf{p}}}{m}$ and $\ka\psi = \psi^*$,
and denoting the 4-component DSF by $\psi =
(\psi_1,\psi_2,\psi_3,\psi_4)^T$ ($\psi_r \in \CC, r=1,\ldots,4$),
 we have the simultaneous conditions a DSF must obey in order for it to be led to an ELKO \cite{osmano}:
\beq 0&=& \mathbb{R}{\rm e}(\psi_1^*\psi_3)=\mathbb{R}{\rm
e}(\psi_2^*\psi_4)\n 0&=&\mathbb{R}{\rm e} (\psi_2^*\psi_3)+
\mathbb{R}{\rm e} (\psi_1^*\psi_4)\n 0&=&
\mathrm{Im}(\psi_1^*\psi_4)-\mathrm{Im}(\psi_2^*\psi_3)-2\mathrm{Im}(\psi_3^*\psi_4)-2\mathrm{Im}(\psi_1^*\psi_2)
\label{partes}.\eeq  In what follows we obtain the extra necessary
and sufficient conditions for each class of DSFs.

As additional conditions on class-(2) Dirac spinor fields, we also
have: \beq \mathbb{R}{\rm
e}(\psi_1^*\psi_4)+\mathrm{Im}(\psi_2^*\psi_3)
&=&0\label{ad2}.\eeq

For the class-(3) of spinor fields, the additional condition was
obtained in \cite{osmano}: \beq
\mathrm{Im}(\psi_1^*\psi_4)-\mathrm{Im}(\psi_2^*\psi_3)-2\mathrm{Im}(\psi_1^*\psi_2)
&=&0\label{ad3}.\eeq

Class-(1) DSFs must obey all the conditions given by
Eqs.(\ref{partes}), (\ref{ad2}), and (\ref{ad3}). Note that if one
relaxes the condition given by Eq.(\ref{ad2}) or Eq.(\ref{ad3}),
DSFs of types-(3) and -(2) are respectively obtained.

Using the decomposition $\psi_j=\psi_{ja}+i\psi_{jb}$ (where
$\psi_{ja}$ = $\mathbb{R}$e($\psi_j$) and $\psi_{jb}$ =
Im($\psi_j$)) it follows that $\mathbb{R}{\rm
e}(\psi_i^*\psi_j)=\psi_{ia}\psi_{ja}+\psi_{ib}\psi_{jb}$ and
$\mathrm{Im}(\psi_i^*\psi_j)=\psi_{ia}\psi_{jb}-\psi_{ib}\psi_{ja}$
for $i,j=1,\ldots,4$. So, in components, the conditions in common
for all types of DSFs are \begin{eqnarray}
\psi_{1a}\psi_{3a}+\psi_{1b}\psi_{3b}&=&0 \label{c1},\\
\psi_{2a}\psi_{4a}+\psi_{2b}\psi_{4b}&=&0 \label{c2},
\end{eqnarray} and the additional conditions for each case are
summarized in Table I below.
\begin{center}
\begin{table}[!h]
\begin{tabular}{|c|c|}
  \hline
  {\bf Class} & {\bf Additional conditions}  \\
  \hline
 \hline (1) & $\psi_{2a}(\psi_{3a}-\psi_{3b})+\psi_{2b}(\psi_{3a}+\psi_{3b}) = 0 = \psi_{3a}\psi_{4b}-\psi_{3b}\psi_{4a}$ \\\hline
  (2) & $\psi_{3a}\psi_{4b}-\psi_{3b}\psi_{4a} = 0 = \psi_{2a}\psi_{3a}+\psi_{2b}\psi_{3b}+\psi_{1a}\psi_{4a}+\psi_{1b}\psi_{4b}$
\\\hline
  (3) & $\psi_{2a}(\psi_{3a}-\psi_{3b})+\psi_{2b}(\psi_{3a}+\psi_{3b})=0$
 and \\
{}&$(\psi_{1a}\psi_{4b}-\psi_{1b}\psi_{4a})-(\psi_{2a}\psi_{3b}-\psi_{2b}\psi_{3a})-2(\psi_{3a}\psi_{4b}-\psi_{3b}\psi_{4a})-$
 $2(\psi_{1a}\psi_{2b}-\psi_{1b}\psi_{2a})=0$ \\
  \hline
\end{tabular}
\caption{Additional conditions, in components, for class (1), (2)
and (3) Dirac spinor fields.}
\end{table}
\end{center}

The explicit mappings obtained above  present the same form of the
instanton Hopf fibration map $S^3\ldots S^7 \rightarrow S^4$
mapping obtained in \cite{jayme}, and can be interpreted as the
geometric meaning of the mass dimension-transmuting operator
obtained in \cite{osmano}, where we obtained mapping between ELKO and Dirac spinor fields.
Some results involving the instanton Hopf fibration can also be seen in this context, e.g, in 
\cite{jal}. In \cite{osmano1} it was shown the reason why the above conditions prevent 
the Hopf fibration to be described by ELKO spinor fields.


\section{The quadratic spinor Lagrangian}
\label{w3}

It is well known that given a spinor-valued 1-form field $\Psi$,
the quadratic spinor Lagrangian (QSL) is given by
\begin{equation}\label{13}
\mathcal{L}_\Psi = 2D\bar{\Psi}\w\g_5 D\Psi =
2\bar\Psi\w\Omega\g_5\w\Psi
 + d[(D\bar\Psi)\w\g_5\Psi + \bar\Psi\w \g_5
D\Psi],
\end{equation}\noi
where $
D\Psi = d\Psi + \omega\w\Psi$ and $D\bar\Psi =
d\bar\Psi + \bar\Psi\w\omega$.
Now, choose the \emph{ansatz}
\begin{equation}\label{35}\Psi = \psi\ot\vartheta,\end{equation}
where $\vartheta$ denotes the orthonormal frame 1-form $\vartheta
= \vartheta^a\ot\g_a = h^a_{\;\,\mu} dx^\mu\ot\g_a$ and $\psi$ is
a spinor field. The action of the spinor covariant exterior
derivative $D$, mapping a spinor-valued 1-form field $\Psi$ into a
spinor-valued 2-form field $D\Psi$ is explicitly given by
$$D\Psi = \vartheta^a\w[\pa^{(s)}\psi\ot\vartheta + \psi\ot(\nabla_{e_a} + (e_a\lrcorner \Theta^c)\w e_c\lrcorner) \vartheta],
$$
where the spin-Dirac operator $\pa^{(s)}$ acting on spinor fields
$\psi$ and the covariant derivative $\nabla_{e_a}$ acting on
Clifford-valued 1-form fields are given respectively by \beq
\pa^{(s)}\psi = \pa_a\psi + \frac{1}{2}\omega_a\psi,\qquad
\nabla_{e_a}\vartheta = \pa_a\vartheta +
\frac{1}{2}[\omega_a,\vartheta], \eeq where $\omega_a =
\omega_{ac}^{b}(e_b\ot\vartheta^c)$.

It is important to remark that the {\it ansatz} 
given by Eq.(\ref{35}) arises in different contexts: in \cite{tn1}
$\psi$ is a Dirac spinor field used to prove the equivalence
between QSL and  the Lagrangians describing General Relativity
(GR), its teleparallel equivalent GR$_\|$, and the M\o ller
Lagrangian; in \cite{tung3} $\psi$ is an auxiliary Majorana spinor
used to prove that gravitation can be described as a SUSY gauge
theory; in \cite{bars,bars1} $\psi$ is an \emph{anticommuting}
Majorana spinor described by Grassmann superspace coordinates,
which generates the spinor supersymmetric conserved current. The
QSL was first proposed in \cite{witten} in the proof of positive
energy theorem.

Up to our knowledge, there are no identities like the
spinor-curvature identities that yield the term linear in
curvature which reduces to the scalar curvature \cite{tung3}. One
of the spinor-curvature identities related which this issue is
given by
\begin{equation}
2D(\bar\psi \xi)\w D(\zeta\psi) = 2(-1)^p\bar\psi
\xi\w\Omega\w(\zeta\psi) + d[\bar\psi\xi\w D(\zeta\psi) - (-1)^p
D(\bar\psi\xi)\w \zeta\psi],
\end{equation}
where now $\xi\in\sec\Lambda^p(T^*M)\ot\cl_{1,3}$ and
$\zeta\in\sec\Lambda(T^*M)\ot\cl_{1,3}$. The scalar curvature
appears in a natural way in the case where $\Psi$ in QSL is a
spinor-valued 1-form field, as suggested in Eq.(\ref{35}).

Substituting the \emph{ansatz} (\ref{35}) in the QSL
(Eq.(\ref{13})), it follows \beq \mathcal{L}_\Psi =
\mathcal{L}(\psi,\vartheta,\omega) &=& 2D(\bar{\psi}\vartheta)\g_5\w D(\vartheta\psi) \nonumber\\
&=&-\bar\psi\psi\,\Omega_{ab}\w\star(\vartheta^a\w\vartheta^b) +
\bar\psi\g_5\psi\,\Omega_{ab}\w\vartheta^a\w\vartheta^b +
d[D(\bar\psi\vartheta)\g_5\psi\vartheta + \bar\psi\vartheta \g_5
D(\vartheta\psi)], \eeq\noi and it is easy to see that when the
spinor field satisfies the normalization conditions
\begin{equation}\label{5}
\bar\psi\psi = 1, \qquad\qquad\bar\psi\g_5\psi =
0,\end{equation}\noi the original QSL can be written as
\begin{equation}\label{6}
\mathcal{L}_\Psi = -\Omega_{ab}\w\star(\vartheta^a\w\vartheta^b) +
d[D(\bar\psi\vartheta)\w\g_5\psi\vartheta + \bar\psi\vartheta
\w\g_5  D(\vartheta\psi)]
\end{equation}\noi
In this way, the DSF $\psi$ enters in the QSL only at the boundary
and does not appear in the equations of motion. In fact, up to the
boundary term, the Lagrangian is given by
\begin{equation}
\mt{L}_\Psi = -\Omega_{ab}\w\star(\vartheta^a\w\vartheta^b),
\end{equation}\noi which is the Einstein-Hilbert Lagrangian.
 Eq.(\ref{6}) shows that the action $S_\Psi = \int\mt{L}_\Psi$ does not depend locally on the Dirac spinor field $\psi$.

Tung and Nester \cite{tn1} asserted that a change on the spinor
field will leave the action $S_\Psi$ unchanged, and then the
spinor field has a six-parameter --- four complex parameters
constrained by Eqs.(\ref{5}) --- local gauge invariance. The
theory also presents a Lorentz gauge freedom related to the
transformations of the orthonormal frame field. They prove that
the spinor field gauge freedom induces a Lorentz transformation on
the orthonormal frame field, and the boundary term has only one
physically independent degree of freedom \cite{tn1}. They also
admit a suitable choice fixing one of the two Lorentz gauges by
tying the DSF to the orthonormal coframe field together. So, the
spinor gauge freedom related to the six parameter DSF $\psi$ is (2
to 1) equivalent to the Lorentz transformations for the associated
orthonormal frame. The choice $d\psi = 0$ clearly implies that
$\psi$ is a constant spinor. However, other choices are possible
where the spinor field $\psi$ is not constant anymore ---
$d\psi\neq 0$. 
%


\section{QSL as the fundament of gravity via the classification of ELKO spinor fields}
\label{ui} Classical spinor fields\footnote{As is well known,
quantum spinor fields are operator valued distributions. It is not
necessary to introduce quantum fields in order to know the
algebraic classification of ELKO spinor fields.} carrying a
$D(1/2,0)\oplus D(0,1/2)$, or $D(1/2,0)$, or $D(0,1/2)$
representation of SL$(2,\CC) \simeq$ Spin$^e_{1,3}$ are sections
of the vector bundle
\begin{equation}
P_{{\rm Spin}^e_{1,3}} (M) \times_\rho \CC^4,\end{equation} \noi
where $\rho$ stands for the $D(1/2,0) \oplus D(0,1/2)$ (or
$D(1/2,0)$ or $D(0,1/2)$) representation of Spin$^e_{1,3}$ in
$\CC^4$. Other important spinor fields, like Weyl spinor fields,
are obtained by imposing some constraints on the sections of
$P_{{\rm Spin}^e_{1,3}}(M) \times_\rho \CC^4$. See, e.g.,
\cite{lou1,lou2} for details. Given a spinor field $\psi$ $\in
\sec\mathbf{P}_{\rm{Spin}_{1,3}^{e}}(M)\times_{\rho}\mathbb{C}^{4}$
the bilinear covariants are the following sections of
${\displaystyle\Lambda
}(T^*M)={\displaystyle\oplus_{p=0}^{4}}$ ${\displaystyle\Lambda^{p}%
}(T^*M)\hookrightarrow C\mathcal{\ell}(M,g)$ \cite{ro1,moro}
\begin{align}
\sigma &  =\bar\psi\psi,\quad\mathbf{J}=J_{\mu}\vartheta%
^{\mu}=\bar\psi\gamma_{\mu}\psi\,\vartheta^{\mu},\quad
\mathbf{S}=S_{\mu\nu}\vartheta^{\mu\nu}=\frac{1}{2}\psi^{\dagger}\gamma
_{0}i\gamma_{\mu\nu}\psi\,\vartheta^{\mu}\wedge\vartheta^{\nu},\nonumber\\
\mathbf{K} &  =\bar\psi i\gamma_{0123}\gamma_{\mu}
\psi\,\vartheta^{\mu},\quad\chi=-\bar\psi\gamma_{0123}%
\psi,\label{fierz}%
\end{align}
with $\sigma,\chi\in\sec \Lambda^0 (T^*M),\,
\mathbf{J,K}\in\sec\Lambda^1 (T^*M)$ and $\mathbf{S}\in\sec
\Lambda^2(T^*M)\hookrightarrow C\mathcal{\ell}(M,g)$. In the
formul\ae\, appearing in Eq. (\ref{fierz}), the set
$\{\gamma_{\mu}\}$ can be thought of as being the Dirac matrices,
but we prefer not to make reference to any kind of representation,
in order to preserve the algebraic character of the theory. When
required, it is possible to use any suitable representation. Also,
$
\{\mathbf{1}_{4},\gamma_{\mu},\gamma_{\mu}\gamma_{\nu},\gamma_{\mu}\gamma
_{\nu}\gamma_{\rho},\gamma_{0}\gamma_{1}\gamma_{2}\gamma_{3}\}$ is
a basis for  $C\mathcal{\ell}(M,g)$, $\mu<\nu<\rho$, and
 $\mathbf{1}_{4}\in$ $\mathbb{C(}4\mathbb{)}$ is the identity matrix.

%
Lounesto spinor field classification --- representation
independent --- is given by the following spinor field classes
\cite{lou1,lou2}, where in the first three classes it is implicit
that $\mathbf{J}$\textbf{, }$\mathbf{K}$\textbf{, }$\mathbf{S}$
$\neq0$:
\begin{enumerate}
\item[(1)] $\sigma\neq0,\;\;\; \chi\neq0$. \item[(2)]
$\sigma\neq0,\;\;\; \chi= 0$. \item[(3)] $\sigma= 0,
\;\;\;\chi\neq0$. \item[(4)] $\sigma= 0 = \chi,
\;\;\;\mathbf{K}\neq0,\;\;\; \mathbf{S}\neq0$. \item[(5)] $\sigma=
0 = \chi, \;\;\;\mathbf{K}= 0, \;\;\;\mathbf{S}\neq0$. \item[(6)]
$\sigma= 0 = \chi, \;\;\; \mathbf{K}\neq0, \;\;\; \mathbf{S} = 0$.
\end{enumerate}

The current density $\mathbf{J}$ is always non-zero. Classes (1),
(2), and (3) are called \textit{Dirac spinor fields} for spin-1/2
particles, and classes (4), (5), and (6) are called, respectively,
\textit{flag-dipole}, \textit{flagpole} and \textit{Weyl spinor
fields}. Majorana and ELKO spinor fields \cite{ro1,alu1,alu2} are
a particular case of a class-(5) spinor field. It is worthwhile to
point out a peculiar feature of spinor fields of class (4), (5),
and (6): although $\mathbf{J}$ is always non-zero, we have
$\mathbf{J}^{2}=-\mathbf{K}^{2}=0$. Although the choices given by
Eq.(\ref{5}) is restricted to class-(2) DSFs, we can explore other
choices for values of $\sigma = \bar\psi\psi$ and $\chi =
\bar\psi\g_5\psi$, and also investigate the QSL from the point of
view of classes (1) and (3) spinor fields.

Now, if instead of class-(2) we consider the class-(3) DSF, in
which case the spinor field satisfies the normalization conditions
\begin{equation}
\sigma = \bar\psi\psi = 0,\qquad \chi = \bar\psi\g_5\psi = 1,
\end{equation}
 then
the original QSL can be written as
\begin{equation}\label{166}
\mathcal{L}_\Psi = -\Omega_{ab}\w(\vartheta^a\w\vartheta^b) +
d[D(\bar\psi\vartheta)\w\g_5\psi\vartheta + \bar\psi\vartheta
\w\g_5  D(\vartheta\psi)].
\end{equation}\noi
The class-(1) DSF $\psi$ enters the QSL only at the boundary, and
consequently it does not appear in the equations of motion. Up to
the boundary term, therefore, the Lagrangian  is given by the
Einstein-Palatini Lagrangian
\begin{equation}
\mt{L}_\Psi = -\Omega_{ab}\w(\vartheta^a\w\vartheta^b).
\end{equation}\noi

It is immediate to see that, by considering a class-(1) DSF,
characterized by the conditions $\sigma\neq 0$ and $\chi\neq 0$,
the most general Holst action, given by
\begin{equation}
 \mt{S}^o_\Psi = \bar\psi\psi\,\int\Omega_{ab}\w\star(\vartheta^a\w\vartheta^b) + \bar\psi\g_5\psi\,\int
\Omega_{ab}\w(\vartheta^a\w\vartheta^b),
\end{equation}
follows naturally \cite{jpe}. In fact, this action comes from the
QSL associated with a class-(1) DSF
\begin{equation} \mt{L}_\Psi = -\bar\psi\psi \Omega_{ab}\w\star(\vt^a\w\vt^b) +\bar\psi\g_5\psi\,\Omega_{ab}\w(\vartheta^a\w\vt^b)
+d[D(\bar\psi\vt)\w\g_5\psi\vt + \bar\psi\vt\w\g_5 D(\vt\psi)].
\end{equation}
The ratio $\sigma/\chi$ --- which
 measures how much Einstein theory departs from a more general
 covariant theory of gravity ---
is exactly the Immirzi parameter, as pointed out by Chou, Tung,
and Yu \cite{imm}. The expectation value of $\sigma/\chi$ also
allows to introduce a renormalization scale upon quantization.

\section{Concluding remarks}
\label{secdis} The main purpose of this paper is to investigate
and discuss the QSL as the fundamental Lagrangian for some of the
current theories for gravity, from the ELKO spinor fields
viewpoint, based also in the previous results in
\cite{ro1,ro3,ro4,ro5,osmano}. It has been shown in \cite{ro1} that ELKO is a type-(5), flagpole spinor field. In addition, 
type-(1) DSFs --- under Lounesto spinor
field classification --- present seven degrees of freedom, and
it can be shown that the mapping from Dirac to ELKO spinor fields
shown in Sec.(\ref{elko}) is a one-to-one correspondence to the
instanton Hopf fibration map $S^3\ldots S^7 \rightarrow S^4$ \cite{jayme}: the 
conditions that the Dirac spinor fields must satisfy in order to be led to ELKO, 
explicitly given by Eqs.(\ref{partes}, \ref{ad2}, \ref{ad3}) are exactly in correspondence 
to the instanton Hopf fibration equation, in the Clifford algebra arena, as shown in \cite{jayme}.
It would suggest the reason why the ELKO spinor
fields satisfy the Klein-Gordon equation, instead of the Dirac
equation. Physically, as ELKO presents mass dimension one
\cite{alu1,alu2,alu3}, while any other type of spin-1/2 spinor
field present mass dimension 3/2, the conditions obtained in
Sec.(\ref{elko}) might introduce the geometric explanation for
this physical open problem.

QSL makes use of a general auxiliary spin-3/2 field that can be
expressed as the tensor product between an auxiliary spinor field
$\psi$ and a Clifford-valued 1-form $\theta$. This auxiliary
spinor field $\psi$ was first introduced by Witten as a convenient
tool in the proof of the positive-energy theorem of Einstein
gravity \cite{witten}. When the QSL is required to yield
Einstein-Hilbert, Einstein-Palatini, and Holst actions, it follows
naturally that the auxiliary spinor-valued 1-form field composing
the QSL takes the form of an ELKO, when we take into account the mapping in Eqs.(\ref{partes},\ref{ad2},\ref{ad3}). Any other choice of spinor field leads, up
to a boundary term, to a null QSL \cite{jpe}. In the light of
Sec.(\ref{elko}), the spinor-valued 1-form field of the QSL must
necessarily be constituted by a    tensor product between an ELKO
spinor field and a Clifford algebra-valued 1-form.

Einstein-Hilbert, Einstein-Palatini, and Holst actions correspond
respectively to  the mapping between ELKO spinor fields and DSFs
of class-(2) (given by Eqs.(\ref{partes}, \ref{ad2})), class-(3)
(given by Eqs.(\ref{partes}, \ref{ad3})), and class-(1) (given by
Eq.(\ref{ad2}, \ref{ad3}, \ref{partes}) DSFs. And, as ELKO spinor
fields can be obtained from the DSFs, via a mapping explicitly
constructed that does not preserves spinor field classes, under
Lounesto classification \cite{osmano}, we conclude that --- in
particular --- the Einstein-Hilbert, Einstein-Palatini, and Holst
actions can be derived from the QSL, as a fundamental Lagrangian
for supergravity, only using ELKO spinor fields\footnote{In \cite{rdar} 
the super-Poincare algebra was obtained in the context of a 3-dimensional Euclidean space.} 

Although the choice $d\psi=0$, and the normalization conditions
$\sigma = \bar\psi\psi = 1$ and $\chi=\bar\psi\g_5\psi = 0$ ---
corresponding to a class-(2) Dirac spinor field --- gives the best
option to prove the equivalence between the QSL and the
Lagrangians associated with general relativity and teleparallel
gravity, they are restrictive if we are interested in more general
analyses. Also, the ELKO spinor field mapping ---
Eqs.(\ref{partes}, \ref{ad2}, \ref{ad3}) --- into classes (2) and
(3) of DSFs can be chosen to give the complete QSL Holst action,
each one corresponding respectively to one of its pieces
$\bar\psi\psi\int\Omega_{ab}\w\star(\vt^a\w\vt^b)$  or
$\bar\psi\g_5\psi\int\Omega_{ab} \w (\vt^a \w \vt^b)$.
Furthermore, the ELKO spinor field mapping (Eqs.(\ref{partes}))
into class-(1) Dirac spinor field  gives alone the complete Holst
action, since in this case $\sigma = \bar\psi\psi \neq 0$ and
$\chi = \bar\psi\g_5\psi\neq 0$.

\section{Acknowledgment}
 The authors are very grateful to Prof. Dharamvir Ahluwalia for pointing out elucidating and enlightening viewpoints.
Rold\~ao da Rocha thanks to Funda\c c\~ao de Amparo \`a Pesquisa
do Estado de S\~ao Paulo (FAPESP) (2008/06483-5) and J. M. Hoff
da Silva thanks to CAPES-Brazil for financial support.

\end{document}